\documentclass[aps,pra,amsmath,amsfonts,amssymb,twocolumn,superscriptaddress,showpacs]{revtex4}

\usepackage{amssymb}
\usepackage{graphicx,psfrag,color}
\usepackage{dcolumn}
\usepackage{bm}
\usepackage{mathbbol}

\begin{document}
\flushbottom

\title{Dynamically Disordered Quantum Walk as a Maximal Entanglement Generator}

\author{Rafael Vieira}
\author{Edgard P. M. Amorim}
\email{eamorim@joinville.udesc.br}
\affiliation{Departamento de F\'isica, Universidade do Estado de
Santa Catarina, 89219-710, Joinville, SC, Brazil}
\author{Gustavo Rigolin}
\email{rigolin@ufscar.br}
\affiliation{Departamento de F\'isica, Universidade Federal de
S\~ao Carlos, 13565-905, S\~ao Carlos, SP, Brazil}

\date{\today}

\begin{abstract}
We show that the entanglement between the internal (spin)
and external (position) degrees of freedom of a qubit in a
random (dynamically disordered) one-dimensional discrete time quantum random walk (QRW)
achieves its maximal possible value asymptotically in the number of steps,
outperforming the entanglement attained by using ordered QRW. The disorder
is modeled by introducing an extra random aspect to QRW,
a classical coin that randomly dictates which quantum coin drives
the system's time evolution. We also show that maximal entanglement is achieved
independently of the initial state of the walker, study the number of steps the system
must move to be within a small fixed neighborhood of its asymptotic limit, and propose two
experiments where these ideas can be tested.
\end{abstract}

\pacs{03.65.Ud, 03.67.Bg, 05.40.Fb}

\maketitle

\textit{Introduction.} 
Imagine we have a qubit, a quantum particle that in
addition to its external degrees of freedom (position and
momentum) has a spin-1/2-like internal one (two level system)
\cite{mudanca1}. 
We assume it evolves in time as follows.
We first apply a unitary operation $C$
(our ``quantum coin'') acting only on the qubit's internal degree of
freedom, leaving it generally in a superposition of spin up and
down. We then apply another unitary operation $S$
that correlates the displacement of the qubit to
its internal degree of freedom.
It moves right if the
spin state at a given site
is up and left otherwise. In this way we entangle the internal and
external degrees of freedom of the system. 
Successive applications of the previous procedure lead to the discrete time
evolution (displacement) of the qubit. 
This is what we call the one-dimensional discrete
time quantum random walk (QRW) \cite{aha93,kem03}.

The key difference between the classical random walk (CRW) \cite{per05ray05} 
and QRW is the superposition
principle of quantum mechanics, a feature that is obviously lacking
in CRW. The application of $C$ followed by the
displacement operator $S$ at each step generates a cat-like state
among all possible positions of the particle, setting the stage for
interference effects to take place. The interference among the
probability amplitudes manifests itself producing a position
probability distribution $P(j)$ drastically different from the
classical one. Indeed, $P(j)$ for the unbiased CRW is always peaked
about the initial position and drops off exponentially with the
square of the distance (Gaussian distribution). Also, its variance $\sigma^2$ is
proportional to the number $n$ of steps (coins flipped). This is the
diffusive behavior. For the unbiased QRW, however, $P(j)$ is roughly
uniform as we move away from the origin, having peaks far from it.
Moreover, depending on the initial spin state we can have one peak
at the left, or at the right, or two symmetrical peaks \cite{kem03},
and $\sigma^2\propto n^2$, a quadratic gain (ballistic behavior) in
the propagation of the particle when compared to CRW. Furthermore,
due to the $SU(2)$ structure of $C$, we have a coin with
three independent parameters while classically there is only 
one.

Both CRW and QRW, in the one or higher dimensional versions, have
many important applications \cite{application}. And the majority of 
studies dealing with QRW assume that
$C$ is the same during all steps of the walk or changes in a 
deterministic way \cite{kem03,she03,eng07,chi09,lov10,car05,bos09}. 
What would happen, though, if noise, disorder, or fluctuations change $C$
from one step to the other? What would happen if
$C$ changes randomly between two possible coins? A naive
guess would suggest that \textit{all} features of QRW may be washed
out by such a process. Indeed, it is known that \textit{some}
typical features of QRW, such as $P(j)$ and $\sigma^2$,
change in such random processes and approach the
classical case \cite{rib04}.
However, so far no systematic numerical and/or analytical studies 
along this line were done for the entanglement content of the walker
and for any initial condition. The only exception is
Ref. \cite{cha12}, which came to our knowledge after finishing this work,
and where for only one initial condition and a particular
type of static and dynamical disorder the behavior of entanglement 
was numerically investigated for a 100-step walk.

Our main goal here is to investigate such extra random aspect on a
quantum random walk (QRW) and analyze whether or not it is detrimental to its
entanglement generation capacity. And our main finding is,
surprisingly, that the opposite from the naive guess occurs when
it comes to entanglement generation using a dynamically 
disordered QRW. We show
that the entanglement, a genuine quantum feature,
between the internal and external degrees of
freedom of the walker is enhanced when $C$ changes
from one step to the other in a truly random way. We also show
that we achieve, asymptotically in the number of steps, a
maximally entangled state. Moreover, we show that this effect is
independent of the initial condition, contrary to standard
entanglement generation schemes that rely critically on the
initial state of the system and never achieve maximal entanglement
\cite{car05}. 
It is worth mentioning that this initial state
independence that we show here has important practical consequences and shows that
the entanglement generation scheme here presented is robust
against imperfections in the preparation of the initial state.

In order to explore these ideas we introduce a walker that
combines the features of both the classical and quantum ones in a
single formalism. It has two random ingredients, one of which is a
classical coin similar to that of CRW. This coin dictates which
quantum coin (the source of position randomness) will be used at each
step of the walk. This is the essence of this walker and the
presence of these two different random aspects, one classical and
another quantum, leads us to call it a random quantum random walk
(RQRW) process. We show in Appendix \ref{ap0} that CRW and QRW are two
particular cases of RQRW. Note that the quantum random aspect manifests 
itself only when we measure the position or spin of the walker 
(measurement postulate of quantum mechanics). The dynamics
is unitary, however, leading some authors to call the ordered case simply
QW instead of QRW.

\textit{Mathematical formalism.} The Hilbert
space of RQRW is $\mathcal{H}=\mathcal{H}_C\otimes \mathcal{H}_P$,
where $\mathcal{H}_C$ is a two-dimensional complex vector space
associated to the spin states
$\{|\!\uparrow\rangle,|\!\downarrow\rangle\}$ and $\mathcal{H}_P$ is
an infinite-dimensional but countable complex Hilbert space spanned by all
integers. Its base is represented by the kets $|j\rangle$, $j\in
\mathbb{Z}$, and they denote the position of the qubit on the
lattice. With this notation we write an arbitrary initial state of
the qubit (walker) as
$
|\Psi(0)\rangle =
\sum_{j}(a(j,0)|\!\uparrow\rangle$ $\otimes|j\rangle+b(j,0)|\!\downarrow\rangle\otimes|j\rangle),
$
with $\sum_j(|a(j,0)|^2+|b(j,0)|^2)=1$ being the normalization
condition and $j$ running over all integers. The time $t$ is discrete
and it denotes the steps of the walker. In a $n$-step process
the time changes from $t=0$ to $t=n$ in increments of one and
the walker's state is
$
|\Psi(n)\rangle = U(n) \cdots
U(1)|\Psi(0)\rangle=\mathcal{T}\prod_{t=1}^{n}U(t)|\Psi(0)\rangle,
$
where $\mathcal{T}$ denotes a time-ordered product, and
\begin{equation}
U(t) = S(C(t)\otimes \mathbb{1}_P).
\label{evolution}
\end{equation}
Here $\mathbb{1}_P$ is the identity operator acting on the space
$\mathcal{H}_P$, $C(t)$ the time-dependent quantum coin,
and $S$ the conditional displacement operator. The operator $S$
moves the qubit at site $j$ to the site $j+1$ if its spin is up and
to the site $j-1$ if its spin is down. Using the present
notation 
$
S=\sum_{j}(|\!\uparrow\rangle\langle\uparrow\!|\otimes|j+1\rangle\langle
j|
+|\!\downarrow\rangle\langle\downarrow\!|\otimes|j-1\rangle\langle
j|).
$

An arbitrary $C(t)$ is given by the most general way of writing an $SU(2)$ unitary
transformation. Up to an irrelevant global phase we have
$
C(t)=c_{\uparrow\uparrow}(t)$ $|\uparrow\rangle\langle\uparrow| +
c_{\uparrow\downarrow}(t) |\uparrow\rangle\langle\downarrow| +
c_{\downarrow\uparrow}(t)|\downarrow\rangle\langle\uparrow| +
c_{\downarrow\downarrow}(t) |\downarrow\rangle\langle\downarrow|,
$
with $c_{\uparrow\uparrow}(t)=\sqrt{q(t)}$,
$c_{\uparrow\downarrow}(t)=\sqrt{1-q(t)}e^{i\theta(t)}$,
$c_{\downarrow\uparrow}(t)=\sqrt{1-q(t)}e^{i\varphi(t)}$, and
$c_{\downarrow\downarrow}(t)=-\sqrt{q(t)}e^{i(\theta(t)+\varphi(t))}$.
Here $0\leq q(t)\leq1$ and $0\leq\theta(t),\varphi(t)\leq 2\pi$. 
The first parameter controls the bias of
$C(t)$. For $q(t)=1/2$ the coin creates an equal superposition of
the spin states when acting on either $|\!\uparrow\rangle$ or
$|\!\downarrow\rangle$ and an unbalanced one for $q(t)\neq 1/2$. The
last two parameters control the relative phase between the two
states in the superposition. Note that we are exploring the full
$SU(2)$ structure of $C(t)$ with its three independent parameters,
which makes it more general than the ones in \cite{rib04}.
Time-dependent walkers were also explored in \cite{bue04}, where
instead of $C$, $S$ was made time-dependent, and in \cite{sha03,ahl11}.

The general time evolution can be obtained applying $U(t)$, Eq.~(\ref{evolution}),
to an arbitrary state at time $t-1$.
This leads to $|\Psi(t)\rangle$ $=$ $U(t)|\Psi(t-1)\rangle$ $=$
$\sum_j(a(j,t)|\!\uparrow\rangle|j\rangle$ $+$ $b(j,t)|\!\downarrow\rangle|j\rangle)$,
where
\begin{eqnarray}
a(j,t) \!&\!=\!&\! c_{\uparrow\uparrow}(t)a(j-1,t-1)+ c_{\uparrow\downarrow}(t)b(j-1,t-1),
\nonumber \\
b(j,t) \!&\!=\!&\! c_{\downarrow\uparrow}(t)a(j+1,t-1)+ c_{\downarrow\downarrow}(t)b(j+1,t-1).
\label{recurrence}
\end{eqnarray}

\begin{figure}[!ht]
\includegraphics[width=8.5cm]{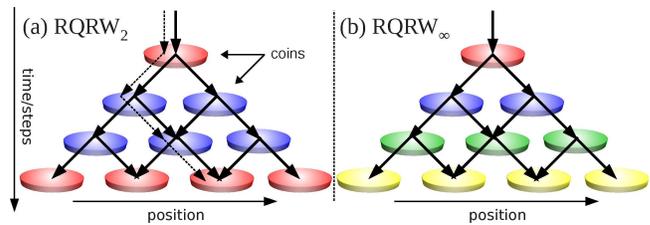}
\caption{\label{fig_esquema}(color online) (a) The dashed line
represents a possible realization of $CRW$, where no superposition
occurs. The solid curves represent probability amplitudes for
$RQRW_2$, where only two $C(t)$ are allowed (red and
blue discs). (b) Schematic view for $RQRW_\infty$, where $C(t)$
are chosen randomly from uniform continuous distributions of
quantum coins (at each step a different color/coin is used). Note that at each
step all coins/colors are the same (dynamical disorder).}
\end{figure}

We will focus here on two types of RQRW (see Fig.
\ref{fig_esquema}). The first one deals with only two quantum coins,
$C_1$ and $C_2$. At each step of the walk the decision to use $C_1$
or $C_2$ is made by the result of a classical coin. If we get heads
at step $t$ we use $C_1$ and if we get tails we use $C_2$. We call this process a
$RQRW_2$, with the subindex denoting that our choices are made
randomly between two quantum coins.

In the second RQRW we have an infinite number of
$C(t)$ to choose at each step. The independent parameters of $C(t)$, namely,
$q(t)$, $\theta(t)$, and $\varphi(t)$, are chosen from continuous uniform distributions
spanning the range of their allowed values. Note that we can have a walk where
either one, or two or all parameters
change at each step. We call such walks $RQRW_\infty$.

\textit{Entanglement.} Since $\rho(t)=|\Psi(t)\rangle\langle\Psi(t)|$ is  pure
we quantify the entanglement between the internal and external
degrees of freedom
by the von Neumann entropy
of the partially reduced state $\rho_C(t)=Tr_P(\rho(t))$ \cite{ben96},
$S_E(\rho(t))=-Tr(\rho_C(t)\log_2\rho_C(t))$, with
$Tr_P(\cdot)$ being the trace over the
position degrees of freedom. $S_E$ is $0$ for separable states and
$1$ for maximally entangled ones. Since
$
\rho_C(t) \!\!=\!\!
\alpha(t) |\!\uparrow\rangle\langle\uparrow\!|\!+\!
\beta(t) |\!\downarrow\rangle\langle\downarrow\!| \!+\!
\gamma(t)
$
$ |\!\uparrow\rangle\langle\downarrow\!|
\!+\! \gamma^*(t) |\!\downarrow\rangle\langle\uparrow\!|,
$
where
$
\alpha(t)=\sum_{j} |a(j,t)|^2,
\beta(t)=\sum_{j} |b(j,t)|^2,
\gamma(t)=\sum_{j} a(j,t)b^*(j,t),
$
and $z^*$ is the complex conjugate of $z$, we have
$
S_E(\rho(t))=-\lambda_+(t)log_2\lambda_{+}(t)-\lambda_{-}(t)log_2\lambda_{-}(t),
$
with
$
\lambda_{\pm}=(1/2{\pm}\sqrt{1/4-\alpha(t)(1-\alpha(t))+|\gamma(t)|^2})
$
being the eigenvalues of $\rho_C(t)$.

\textit{Results.} We start studying two typical representatives of RQRW.
The first one is $RQRW_2$ with $C_1$ being the Hadamard ($H$) coin ($q(t)=1/2$,
$\theta(t)=\varphi(t)=0$) and $C_2$ the Fourier/Kempe ($F$) coin ($q(t)=1/2$,
$\theta(t)=\varphi(t)=\pi/2$).  Note that the latter coin introduces
a $\pi/2$ relative phase between $|\!\uparrow\rangle$ and $|\!\downarrow\rangle$.
The other walk is $RQRW_\infty$, where at each step
the values of $q(t), \theta(t)$ and $\varphi(t)$ are chosen randomly from three distinct
continuous uniform distributions.
\begin{figure}[!ht]
\includegraphics[width=8cm]{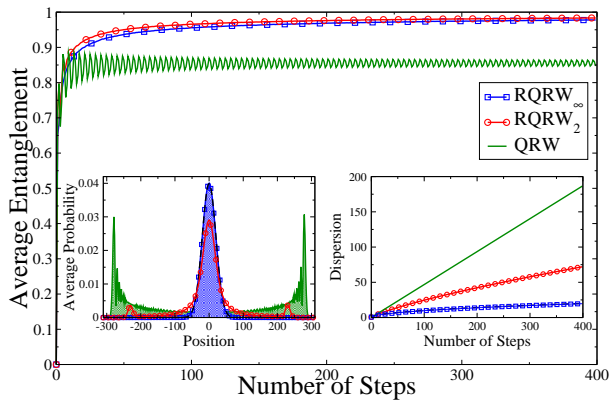}
\caption{\label{figB}(color online) $\langle S_E \rangle$ was
computed averaging over 16,384 initial conditions of the
form,
$|\Psi(0)\rangle\hspace{-.1cm}=\hspace{-.1cm}(\cos\alpha_s|\hspace{-.1cm}\uparrow\rangle+e^{i\beta_s}
\hspace{-.1cm}\sin\alpha_s|\hspace{-.1cm}\downarrow\rangle) \otimes
(\cos\alpha_p|-1\rangle+e^{i\beta_p}\hspace{-.1cm}\sin\alpha_p|1\rangle)$,
with $\alpha_{s,p}\in [0,\pi]$ and $\beta_{s,p}\in [0,2\pi]$. The
first realization used the initial condition
$(\alpha_{s},\beta_{s},\alpha_{p},\beta_p)=(0,0,0,0)$ and the
subsequent ones all quadruples of points in independent increments of $0.4$ until
$\alpha_{s,p} = \pi$ and $\beta_{s,p} = 2\pi$. We worked with a $400$-step
walk. The blue/square curve gives $RQRW_\infty$, the red/circle
one $RQRW_2$, and the green/solid one the Hadamard $QRW$. The left
inset shows the average position probability distributions ($\langle P(j)
\rangle$)  after $400$ steps and the right one  
the average dispersions ($\langle\sqrt{\sigma^2}\rangle$). 
The black/dashed curves represent the expected results for CRW starting at the origin, 
and they overlap with the curves for
$RQRW_\infty$, which is more localized
than $RQRW_2$. 
The two spikes on $\langle P(j) \rangle$ for $RQRW_2$
is due to the fact that it is built on the Hadamard and Fourier
coins, where these spikes are a common trend.}
\end{figure}

In order to investigate the dependence of the asymptotic behavior of $S_E$
on initial conditions, we run several thousands
numerical experiments, each of which with a different initial condition. Each realization of the
walk gives at step $t$ a value for $S_E$ and in Fig. \ref{figB} we show the
average values of $S_E$ over all realizations at each step $t$.

As can be seen from Fig. \ref{figB}, the average entanglement $\langle S_E \rangle$
approaches the maximal value possible
($S_E=1$) for both RQRW cases after a few hundreds steps. 
For comparison, we show the usual QRW with a Hadamard
coin, where clearly
$\langle S_E \rangle \neq 1$
asymptotically.
Indeed, for the ordered case the asymptotic value of $S_E$ is
highly sensitive to the initial conditions and the set of initial states giving high values of
$S_E$ is not dense. An important example
is the Hadamard walk, where it can be shown \cite{car05} that
the asymptotic values of $S_E$ continuously oscillate between
$S_E=0.661$ and $S_E=0.979$ as we cover a
set of initial conditions similar to the ones in  Fig. \ref{figB}.

\begin{figure}[!ht]
\includegraphics[width=8cm]{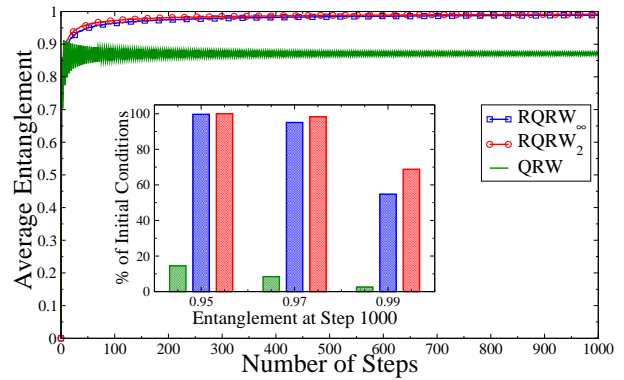}
\caption{\label{figC}(color online) $\langle S_E \rangle$
was got averaging over
2,016 localized initial conditions given as
$|\Psi(0)\rangle=(\cos\alpha_s|\hspace{-.1cm}\uparrow\rangle+e^{i\beta_s}
\hspace{-.1cm}\sin\alpha_s|\hspace{-.1cm}\downarrow\rangle)$ $\otimes
|0\rangle$,
where $\alpha_{s}\in [0,\pi]$ and $\beta_{s}\in [0,2\pi]$. The first realization used the initial
condition $(\alpha_{s},\beta_{s})=(0,0)$ and the next ones
all pairs of points in independent increments of $0.1$ until 
$\alpha_s=\pi$ and $\beta_s=2\pi$. We worked with a $1000$-step walk.
The inset shows the rate of initial conditions leading to $S_E$ greater than
$0.95$, $0.97$, and $0.99$ at step $1000$. Note that for RQRW (middle/blue and right/red bars)
almost $100\%$ of the initial conditions lead to $S_E>0.97$ while for QRW
(left/green bar) this occurs for less than $10\%$.}
\end{figure}

To gain further insights into the asymptotic limit of $S_E$ we run another set of
numerical experiments for the three walks described in Fig.~\ref{figB}, but now going up to
1000 steps and also counting the number of initial conditions leading to high values of
$S_E$. Looking at Fig.~\ref{figC} it is clear that $S_E \rightarrow 1$ for $RQRW_\infty$ and $RQRW_2$
while the Hadamard $QRW$ asymptotic entanglement is highly sensitive to the initial conditions.

Now, since $S_E$ is bounded from above
by one, $\langle S_E\rangle \rightarrow 1$ implies that for RQRW
the set of initial states in which $S_E\rightarrow 1$ asymptotically
is dense. In other words, this suggests that in the asymptotic limit $S_E\rightarrow 1$
for any initial condition. The justification of the last assertion is 
given by the following theorem.
\begin{quote}
\textit{Theorem.} In the asymptotic limit and for any initial
condition, $S_E\rightarrow 1$ if the quantum coin acting on the walker at
each step is a random $SU(2)$ unitary operator.
\end{quote}

Here we outline the main ideas leading to the proof and the details are given in Appendix \ref{apA}.
In the long time regime $\rho_C(t+1)=\rho_C(t)+\mathcal{O}(t^{-1/4})$. 
Thus, $\rho_C(t+1)$ $=$ $\rho_C(t)$ for $t$ $\rightarrow$ $\infty$. 
In terms of its coefficients $\alpha(t+1)=\alpha(t)$ and $\gamma(t+1)=\gamma(t)$.
This, plus the time evolution of $\alpha(t)$ and $\gamma(t)$, 
that can be computed with Eq.~(\ref{recurrence}), leads to
$\mathrm{Re}$ $[(c_{\uparrow\uparrow}(t+2)/c_{\uparrow\downarrow}(t+2)$ $-$
$c_{\uparrow\uparrow}(t+1)/c_{\uparrow\downarrow}(t+1))$ $\gamma(t)]$ $=$ $0$.
Note that for constant coins, this equality is trivially satisfied. But for 
time dependent random ones, the term inside the parenthesis is a random complex number 
$z(t)=x(t) + iy(t)$, with $x(t)$ and $y(t)$ random reals. Hence 
$
x(t)\mathrm{Re} [\gamma(t)] - y(t)\mathrm{Im} [\gamma(t)]= 0.
$
Repeating this argument for a subsequent time leads to
$
x(t+1)\mathrm{Re} [\gamma(t)] - y(t+1)\mathrm{Im} [\gamma(t)]= 0.
$
These expressions form a homogeneous system of linear equations
on the variables $\mathrm{Re} [\gamma(t)]$ and $\mathrm{Im} [\gamma(t)]$. 
A non-trivial solution exists if the determinant of its coefficients is zero. But this will
almost surely not happen since $x(t),y(t),x(t+1)$, and $y(t+1)$ are four independent random numbers.
Thus $\gamma(t)=0$ and $\alpha(t)=1/2$, since
the dynamics and the asymptotic condition give $\alpha(t)= 1/2+ 
\mathrm{Re}[\gamma(t)c_{\uparrow\uparrow}(t+1)/c_{\uparrow\downarrow}(t+1)]$.
These values for $\gamma(t)$ and $\alpha(t)$ gives $S_E=1$,
a maximally entangled state.

In Appendix \ref{apB} we investigate numerically other RQRW, some of them not covered by the theorem,
and how much disorder we must have to achieve $S_E\rightarrow 1$ asymptotically. We  show
that weak disorder is sufficient to generate highly entangled states for arbitrary initial conditions in a
variety of RQRW and give further details about the probability distribution of the walker and its dispersion properties.
Finally, we also investigate how fast highly non-local initial conditions (Gaussian distributions)
approach the asymptotic limit $S_E \rightarrow 1$.

\textit{Experimental implementation.} Current technology allows one to implement in at
least two ways the previous walks. The first one is based on passive optical
elements, such as quarter (QWP) and half (HWP) wave plates and
polarizing beam-splitters (PBS), plus a fast-switching electro-optical modulator (EOM)
\cite{sch10}, where the internal degree of freedom of the walker
is the polarization of a photon and the position/external one is mapped
to different arrival times of the photon at the photodetector (time bins)
 \cite{experiment1}.

The second way also uses photons as walkers but it
is based on integrated photonics, where a disordered
walk is built on integrated waveguide circuits, providing
perfect phase stability. By using state-of-the-art
femtosecond laser writing techniques, the authors in \cite{cre13}
were able to wrought an array of interferometers in a glass
that reproduces the dynamics of RQRW \cite{experiment2}.

To test the ideas here presented we need to measure the entanglement of
the walker, which is obtained if we know the coin state $\rho_C(t)$.
But $\rho_C(t)$ is determined by slightly changing the two schemes
outlined above. Indeed, since a general photon polarization state is written
as $\rho_C(t)=\mathbb{1}_C+\sum_{j=1}^{3}r_j\sigma_j$, with $\sigma_j$ being
Pauli matrices, we can determine $\rho_C(t)$ if we measure $r_j$.
But this is achieved by measuring the average polarization of the photon
in the vertical/horizontal axis ($r_3$), in the $\pm 45^o$ axis ($r_1$), and the average
right/left circular polarization ($r_2$) \cite{per02}. These measurements can
be easily implemented by properly arranging a HWP and QWP before the photon
passes a PBS with photodetectors at each one of its arms. Note that the raw data
are related to $\rho(t)$ and we need to trace out its position degrees
of freedom (post-processing measurement) to get $\rho_C(t)$. In Appendix \ref{apC} we
show that just a few steps are enough to have different predictions for the
behavior of $S_E$ if we work with either $RQRW_2$ or QRW.

\textit{Summary.} We defined the random quantum random walk (RQRW),
a discrete time quantum random walk scheme whose unitary
evolution at each step is chosen randomly using a two-sided (or
infinitely-sided) classical coin. We showed that both the usual
classical and quantum random walks are particular cases of RQRW.
We then studied its entanglement generation capacity.
We showed that RQRW creates
maximally entangled sates in the asymptotic limit for several types of 
dynamical disorder (random time evolution), contrary to the
ordered QRW. Furthermore, and surprisingly,  
we proved that RQRW entanglement creation capabilities
are independent of the initial condition of the walker, another
property in contrast to ordered QRW.

Finally, we would like to point out that our findings naturally
lead to new important questions. For example, what is the
interplay between order/disorder and entanglement creation for
two- or three-dimensional walkers? What would happen to the entanglement
for static disorder \cite{cha12,sch10,cre13}? Can the previous results be
adapted to the case of
two or more \cite{bos09}
walkers to improve the creation of bipartite and
multipartite entanglement, respectively, only among the internal
degrees of freedom? We believe investigations along these lines
may bring other unexpected and intriguing results and foster the development of
new entanglement generation protocols.

\begin{acknowledgments}
The authors thank the anonymous referee for many insightful 
suggestions that improved the presentation of this manuscript.
RV thanks CAPES (Brazilian Agency for the Improvement of Personnel of Higher Education)
for funding. GR thanks the Brazilian agencies CNPq
(National Council for Scientific and Technological Development) and
FAPESP (State of S\~ao Paulo Research Foundation) for funding and
CNPq/FAPESP for financial support through the National Institute of
Science and Technology for Quantum Information.
\end{acknowledgments}

\appendix

\section{Proof that QRW and CRW are particular cases of RQRW}
\label{ap0}

Consider a one-dimensional lattice of
regularly spaced points, where each point corresponds to the
positions a classical particle can be found. Let us
assume this particle moves right  or left
according to the result of a coin tossing
game. The probability of obtaining heads (moves right) is $p$ and
of getting tails (moves left) is $1-p$. This process is known as
the one-dimensional discrete time classical random walk (CRW)
and we call it unbiased or symmetric
if we deal with a fair coin ($p=1/2$) and biased otherwise.

Looking at $C(t)$, Eq.~(\ref{qcoin}), we see that it  changes at each step by how $q(t)$, $\theta(t)$, and
$\varphi(t)$ change with time. If they are
constant in time we recover QRW. Moreover, assume the system's
initial state is $|\!\phi\rangle\otimes|0\rangle$, with
$|\phi\rangle$ an arbitrary spin state and let the initial coin be
such that $C(1)|\phi\rangle=|\!\uparrow\rangle$. This can always be
achieved since $C(t)$ is an arbitrary $SU(2)$ rotation. 
Since $U(t) = S(C(t)\otimes \mathbb{1}_P)$ we see that after the first step the particle moves
right. Now, for $t\geq 2$ let $C(t)$ be chosen between two
choices according to the result of a classical coin tossing in the
following way. If one gets heads $C(t)$ is chosen such that the
particle moves right and if one gets tails it is chosen such
that the particle moves left. The first case is achieved by
choosing $C(t)=\mathbb{1}_C$ ($\sigma_1$) if the spin state is
$|\!\uparrow\rangle$ ($|\!\downarrow\rangle$) and the second one by
choosing $C(t)=\sigma_1$ ($\mathbb{1}_C$) if the spin state is
$|\!\uparrow\rangle$ ($|\!\downarrow\rangle$), where $\sigma_1$ is
the spin flip operator ($q(t)=\theta(t)=\varphi(t)=0$). It is
not difficult to see that this is an exact simulation of CRW and a
proof that it is a particular case of RQRW.

\section{Proof of Theorem}
\label{apA}

\subsection{The proof}

We want to prove the following theorem:
\begin{quote}
\textit{Theorem.} In the asymptotic limit and for any initial condition, $S_E\rightarrow 1$ if the 
quantum coin acting on the walker at each step is a random $SU(2)$ unitary operator.
\end{quote}

Before we start, we must clarify what we mean by ``the asymptotic limit''. First, the 
asymptotic limit is associated to the long time behavior of the reduced density matrix $\rho_C(t)=Tr_P(\rho(t))$ 
describing the internal degrees of freedom of the system.
Here $\rho(t)=|\Psi(t)\rangle\langle\Psi(t)|$ and
$|\Psi(t)\rangle$ $=$ $\sum_j(a(j,t)|\!\uparrow\rangle|j\rangle$ $+$ $b(j,t)|\!\downarrow\rangle|j\rangle)$.
This asymptotic condition was analytically proved for QRW in the second reference of \cite{car05} and in Sec. \ref{asympt} we
numerically show that this is also true for RQRW. More specifically, we show that in the long time regime
$\rho_C(t+1)=\rho_C(t)+\mathcal{O}(t^{-1/4})$, where $\mathcal{O}(t^{-1/4})$ is at most of order $t^{-1/4}$. Therefore,
when we invoke the asymptotic limit, it is implied that we are in the limit $t\rightarrow \infty$ and 
slightly abuse notation by writing $\rho_C(t+1)=\rho_C(t)$ from the start.
Note that all steps in the proof could be carried out by using $\rho_C(t+1)=\rho_C(t)+\mathcal{O}(t^{-1/4})$ 
and taking the limit at the last step of the proof.  This would introduce in all expressions below an 
$\mathcal{O}(t^{-1/4})$ term which would vanish when $t\rightarrow \infty$.

Let us start the proof by writing
\begin{displaymath}
\rho_C(t) =
\left(
\begin{array}{cc}
\alpha(t) & \gamma(t)\\
\gamma^*(t)&  \beta(t)
\end{array}
\right),
\end{displaymath}
where $\alpha(t)=\sum_{j} |a(j,t)|^2$, $\beta(t)=\sum_{j} |b(j,t)|^2$, and
$\gamma(t)=\sum_j a(j,t)b^*(j,t)$. With this notation, the asymptotic limit as 
discussed above implies $\alpha(t+1)=\alpha(t)$,
$\beta(t+1)=\beta(t)$, and $\gamma(t+1)=\gamma(t)$. 

For \textit{any initial condition}, the time evolution of the system for a given RQRW is given by
\begin{eqnarray}
a(j,t+1) \!&\!=\!&\! c_{\uparrow\uparrow}(t+1)a(j-1,t)+ c_{\uparrow\downarrow}(t+1)b(j-1,t),
\nonumber \\
b(j,t+1) \!&\!=\!&\! c_{\downarrow\uparrow}(t+1)a(j+1,t)+ c_{\downarrow\downarrow}(t+1)b(j+1,t),
\nonumber \\
\label{Arecurrence}
\end{eqnarray}
where different RQRW's are obtained changing the way
$c_{jk}(t)$, $j,k=\uparrow,\downarrow$, evolves with time. 

Using Eq.~(\ref{Arecurrence}) we have
\begin{eqnarray}
\alpha(t+1)&=& \sum_{j} a(j,t+1)a^*(j,t+1)\nonumber \\
&=& |c_{\uparrow\uparrow}(t+1)|^2 \alpha(t) + |c_{\uparrow\downarrow}(t+1)|^2 \beta(t) \nonumber \\
&&+ 2\mathrm{Re}[c_{\uparrow\uparrow}(t+1)c^*_{\uparrow\downarrow}(t+1)\gamma(t)], 
\label{Aeq}
\end{eqnarray}
where $\mathrm{Re}[z]$ is the real part of the number $z$. Now, employing the unitarity
of the coin ($|c_{\uparrow\uparrow}(t+1)|^2+ |c_{\uparrow\downarrow}(t+1)|^2$ $=$ $1$) and the
normalization condition ($\alpha(t)+\beta(t)=1$) we have
\begin{eqnarray}
\alpha(t+1)&=& |c_{\uparrow\downarrow}(t+1)|^2  + (1 - 2 |c_{\uparrow\downarrow}(t+1)|^2) \alpha(t) \nonumber \\
&&+ 2\mathrm{Re}[c_{\uparrow\uparrow}(t+1)c^*_{\uparrow\downarrow}(t+1)\gamma(t)]. 
\end{eqnarray}
Finally, using the assumption that we are in the asymptotic limit ($\alpha(t+1)=\alpha(t)$) we obtain
\begin{equation}
\alpha(t)= \frac{1}{2}+ \mathrm{Re}\left[\frac{c_{\uparrow\uparrow}(t+1)}{c_{\uparrow\downarrow}(t+1)}\gamma(t)\right].
\label{Aeqfim} 
\end{equation}

Similarly, starting with $\beta(t+1)$ in Eq.~(\ref{Aeq}) we get
\begin{equation}
\beta(t)= \frac{1}{2}+ \mathrm{Re}\left[\frac{c^*_{\downarrow\downarrow}(t+1)}{c^*_{\downarrow\uparrow}(t+1)}\gamma(t)\right].
\label{Beqfim} 
\end{equation} 

Invoking again that we are in the asymptotic limit, $\alpha(t+1)=\alpha(t)$, we have after
inserting Eq.~(\ref{Aeqfim}) in the previous relation,
\begin{equation}
\mathrm{Re}\left[\left(\frac{c_{\uparrow\uparrow}(t+2)}{c_{\uparrow\downarrow}(t+2)}-
\frac{c_{\uparrow\uparrow}(t+1)}{c_{\uparrow\downarrow}(t+1)}\right)\gamma(t)\right]=0,
\label{apA1}
\end{equation}
where we have used that $\gamma(t+1)=\gamma(t)$ to arrive at the last equality. 
Since we are dealing with random $SU(2)$ unitary 
matrices, the term inside the parenthesis above is a random complex number 
$z(t)=x(t) + iy(t)$, with $x(t)$ and $y(t)$ random reals. Writing $\gamma(t)=\mathrm{Re} [\gamma(t)] + i \mathrm{Im}[\gamma(t)]$, 
Eq.~(\ref{apA1}) implies
\begin{equation}
x(t)\mathrm{Re} [\gamma(t)] - y(t)\mathrm{Im} [\gamma(t)]= 0.
\label{sistema1}
\end{equation}

We can repeat the previous argument starting with $\alpha(t+2)=\alpha(t+1)$. Using the asymptotic
assumption that $\gamma(t+2)=\gamma(t+1)=\gamma(t)$ leads to
\begin{equation}
x(t+1)\mathrm{Re} [\gamma(t)] - y(t+1)\mathrm{Im} [\gamma(t)]= 0.
\label{sistema2}
\end{equation}

Now, Eqs.~(\ref{sistema1}) and (\ref{sistema2}) constitute a homogeneous system of linear equations
on the variables $\mathrm{Re} [\gamma(t)]$ and $\mathrm{Im} [\gamma(t)]$. We can only achieve a non-trivial solution
if the determinant of its coefficients is zero, i.e., if $x(t)y(t+1)=y(t)x(t+1)$. 
But this will almost surely not happen since $x(t),y(t),x(t+1)$, and $y(t+1)$ are four independent random real numbers and
we have $x(t)y(t+1)\neq y(t)x(t+1)$.
Hence, Eqs.~(\ref{sistema1}) and (\ref{sistema2}) imply
\begin{equation}
\mathrm{Re} [\gamma(t)] = \mathrm{Im} [\gamma(t)]= 0 \rightarrow \gamma(t)=0,
\label{sol1}
\end{equation}
and consequently (see Eqs.~(\ref{Aeqfim}) and (\ref{Beqfim}))
\begin{equation}
\alpha(t) = \beta(t)= 1/2.
\label{sol2}
\end{equation}
Finally, inserting the asymptotic values of $\alpha(t)$, $\beta(t)$, and $\gamma(t)$ just computed
into the expression for $\rho_C(t)$ we get
\begin{displaymath}
\rho_C(t) =
\left(
\begin{array}{cc}
1/2 & 0\\
0&  1/2
\end{array}
\right),
\end{displaymath}
which immediately leads to the maximal allowed value for the
entanglement, $S_E=-Tr(\rho_C(t)\log_2\rho_C(t))=1. \square$


\textit{Remark 1:} The previous proof also applies whenever we have a quantum coin $C(t)$ 
with at least $\theta(t)$ random; and also 
with only random $q(t)$ and with $\theta(t)$ not zero or
a multiple of $\pi/2$. To see that, let us rewrite here the quantum coin,
\begin{equation}
C(t)=
\left(
\begin{array}{cc}
c_{\uparrow\uparrow}(t) & c_{\uparrow\downarrow}(t) \\
c_{\downarrow\uparrow}(t) & c_{\downarrow\downarrow}(t)
\end{array}
\right),
\label{qcoin}
\end{equation}
with
$c_{\uparrow\uparrow}(t)=\sqrt{q(t)}$,
$c_{\uparrow\downarrow}(t)=\sqrt{1-q(t)}e^{i\theta(t)}$,
$c_{\downarrow\uparrow}(t)=\sqrt{1-q(t)}e^{i\varphi(t)}$, and
$c_{\downarrow\downarrow}(t)=-\sqrt{q(t)}e^{i(\theta(t)+\varphi(t))}$.
Here $0\leq q(t)\leq 1$, $0\leq\theta(t)\leq 2\pi$, and
$0\leq\varphi(t)\leq2\pi$ \cite{footnote3}.

If we take the case where at least $\theta(t)$ changes 
randomly with time, with $\varphi(t)$ or $q(t)$ changing or not, 
we see that 
$c_{\uparrow\uparrow}(t)/c_{\uparrow\downarrow}(t)$ $=$ $\sqrt{q(t)/(1-q(t))}e^{-i\theta(t)}$ is a 
random complex number due to the randomness of $\theta(t)$. (Note that arguments based on the last
equation is not valid whenever we have a fixed $q(t)=0$ or $q(t)=1$.) Therefore, Eq.~(\ref{Aeqfim}) has
the same properties as in the original proof and implies Eqs.~(\ref{apA1}), (\ref{sistema1}) and
(\ref{sistema2}). The same argument holds for Eq.~(\ref{Beqfim}).
In other words, the whole proof follows if we guarantee in Eq.~(\ref{Aeqfim}) or
in Eq.~(\ref{Beqfim}) that we have random complex numbers multiplying $\gamma(t)$ at each step. 
By this simple argument we have proved that in the asymptotic limit 
$S_E\rightarrow 1$ for four cases: 
all parameters randomly changing with time (the original proof), $\theta(t)$ and $\varphi(t)$ random with $q(t)$ fixed, 
$\theta(t)$ and $q(t)$ random
with $\varphi(t)$ fixed, and $\theta(t)$ random with both $\varphi(t)$ and $q(t)$ fixed. 

For $q(t)$ random and $\theta(t)$ fixed but not zero or a multiple of $\pi/2$ we have
$c_{\uparrow\uparrow}(t)/c_{\uparrow\downarrow}(t)=\sqrt{q(t)/(1-q(t))}e^{-i\theta(t)}$, which due
to the randomness of $q(t)$ is a random complex number and the proof follows. Note that if $\theta(t)=0$,
$\theta(t)=\pi/2$, or $\theta(t)=\pi$,
the proof cannot be carried out to its completion. In the first and third cases $c_{\uparrow\uparrow}(t)/c_{\uparrow\downarrow}(t)$
is a random real and following the steps of the proof leads only to $\mathrm{Re} [\gamma(t)] = 0$; nothing can
be said about the imaginary part of $\gamma(t)$. And in the second case $c_{\uparrow\uparrow}(t)/c_{\uparrow\downarrow}(t)$
is a random pure imaginary leading to  $\mathrm{Im} [\gamma(t)] = 0$, while nothing can be said about the
real part. 

\textit{Remark 2:} The structure of the proof does not allow us to reach any conclusion when only $\varphi(t)$ changes. 
Noting that 
$c_{\uparrow\uparrow}(t)/c_{\uparrow\downarrow}(t)=\sqrt{q(t)/(1-q(t))}e^{-i\theta(t)}$ and
$c_{\downarrow\downarrow}(t)/c_{\downarrow\uparrow}(t)=$ $-\sqrt{q(t)/(1-q(t))}e^{i\theta(t)}$, we see that
they are both complex constants if only $\varphi(t)$ changes. Thus, since we do not have random complex numbers
multiplying $\gamma(t)$ in either Eq.~(\ref{Aeqfim}) or (\ref{Beqfim}), the proof cannot follow. 

\textit{Remark 3:} Although the proof here cannot be extended to some particular cases that do not explore the full
$SU(2)$ structure of the coin $C(t)$, numerical simulations (see the main text and Sec. \ref{apB}) 
suggest that if we have
at least one of the three independent parameters of $C(t)$ random, we obtain $S_E\rightarrow 1$ asymptotically.
Also, we found no analytical proof that the binary randomness of the balanced or unbalanced $RQRW_2$ 
leads to maximal entanglement asymptotically. 
However, extensive numerical analysis showed that this is true (see main text and Sec. \ref{apB}). 

\textit{Remark 4:} Finally, the proof does not tell us when the system will approach the asymptotic limit.
It may happen for a few hundreds steps or we may need several thousands or more steps. As we show in
Sec. \ref{apB}, the more delocalized in position the initial condition the more steps we need. 

\subsection{Numerical proof of the asymptotic assumption}
\label{asympt}

The previous theorem has two assumptions that led to the proof of its thesis.
The first one is related to the dynamical property of
the walker we are dealing with: we have random $SU(2)$ unitary
coins at each time step.  Without this random feature, i.e., 
if we were dealing with a fixed coin, Eq.~(\ref{apA1}) 
is trivially satisfied and the proof does not follow.
Indeed, for a fixed coin we have $c_{jk}(t)=c_{jk}(t+1)$, with
$j,k = \uparrow,\downarrow$. This makes the term inside
the parenthesis of Eq.~(\ref{apA1}) zero and nothing can be said
about $\gamma(t)$.

The second ingredient is the asymptotic assumption, namely,
$\lim_{t\rightarrow \infty}[\rho_C(t+1)-\rho_C(t)]=0$, a sufficient (but not necessary) 
condition for  $S_E(\rho(t+1))=S_E(\rho(t))$. It is a sufficient
condition for the asymptotic behavior of the entanglement since $S_E(\rho(t))$ 
is a function of the coefficients of $\rho_C(t)$ (see the main text).

For the Hadamard QRW it was analytically shown that the asymptotic assumption is
a consequence of the dynamics of the walker (see second reference of \cite{car05}). 
Our goal here is to provide a numerical proof that the same
fact holds for RQRW. Moreover, we also investigate the rate at which
the asymptotic limit is achieved for both QRW and RQRW. We find
the interesting fact that the two rates obey a power law with different exponents.

In order to quantify the rate at which the asymptotic limit is approached we
compute the trace distance \cite{nie2000} between two adjacent states in time,
\begin{equation}
D(t) = \frac{1}{2}\text{Tr}\left(\left|\rho_C(t) - \rho_C(t-1)\right|\right),
\label{td}
\end{equation}
where  $|A| = \sqrt{A^\dagger A}$.  For a qubit it is not difficult to show
that the trace distance $D(t)$ is equal to the Ky Fan 1-norm (largest singular value of 
$\rho_C(t) - \rho_C(t-1))$ and that the Frobenius norm of $\rho_C(t) - \rho_C(t-1)$ 
equals $\sqrt{2}D(t)$. In other words, the results we obtain in what follows are 
quite independent of the norm we choose to work with. 

A direct computation gives
\begin{equation}
D(t) = \frac{1}{2}\sqrt{(\Delta r_1(t))^2+(\Delta r_2(t))^2+(\Delta r_3(t))^2},
\label{tdqubit}
\end{equation}
where $\Delta r_j(t) = \text{Tr}(\rho_C(t)\sigma_j)-\text{Tr}(\rho_C(t-1)\sigma_j)$,
with $\sigma_j$ being the Pauli matrices.

\begin{figure}[!ht]
\includegraphics[width=8cm]{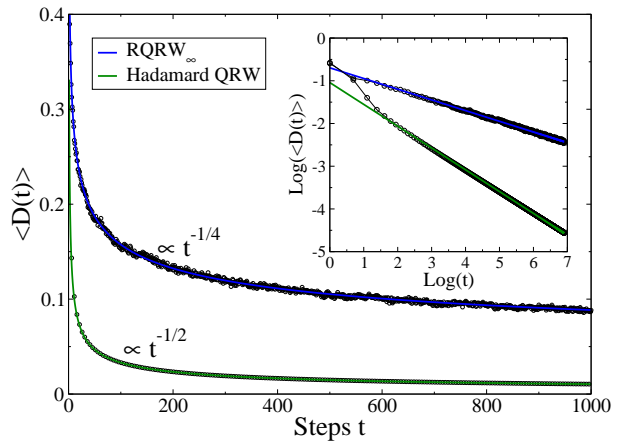}
\caption{\label{figdistancia}(color online)
The curves were plotted averaging 
over 1000 random initial conditions of the following form,
$|\Psi(0)\rangle\hspace{-.1cm}=\hspace{-.1cm}(\cos\alpha_s|\hspace{-.1cm}\uparrow\rangle+e^{i\beta_s}
\hspace{-.1cm}\sin\alpha_s|\hspace{-.1cm}\downarrow\rangle) \otimes 
(\cos\alpha_p|-1\rangle+e^{i\beta_p}\hspace{-.1cm}\sin\alpha_p|1\rangle)$, 
where $\alpha_{s,p}\in [0,\pi]$ and $\beta_{s,p}\in [0,2\pi]$.  We worked with a $1000$-step walk.
The lower curve is the Hadamard QRW and the upper curve $RQRW_{\infty}$ 
with all its three parameters randomly chosen at each step from the following 
uniform distributions: $q(t)\in [0,1], \theta(t)\in [0,\pi]$, and $\varphi(t)\in [0,2\pi]$. 
The small circles are the simulated 
$\langle D(t)\rangle$ and the solid lines fitted curves to the data.
The inset shows the log-log plot of the main graph. A $95\%$ confidence
level linear fitting gives, as expected, $-0.2504\pm 0.0005$ for the slope of
the $RQRW_\infty$ line and $-0.5102 \pm 0.0007$ for the slope of the 
Hadamard QRW line. If we drop the initial points to implement the fitting 
and focus on the asymptotic ones, 
the slopes approach even more the respective values $-1/4$ and $-1/2$.}
\end{figure}

The first series of numerical experiments we implemented is shown in Fig.~\ref{figdistancia}.
We worked with the Hadamard QRW and $RQRW_{\infty}$ with random $q(t)$, $\theta(t)$,
and $\varphi(t)$. We computed $D(t)$ for hundreds of random initial
conditions and plotted the average trace distance. 
As can be seen from the fitted curves, both QRW and $RQRW_{\infty}$
approach the asymptotic limit ($\langle D(t)\rangle \rightarrow 0$) with a power law. 
For QRW a $95\%$ confidence level non-linear fitting gave 
$D(t) = (0.331217\pm 0.000006)t^{-1/2}$ and  $D(t) = (0.4977 \pm 0.0003) t^{-1/4}$ for 
$RQRW_{\infty}$.

\begin{figure}[!ht]
\includegraphics[width=8cm]{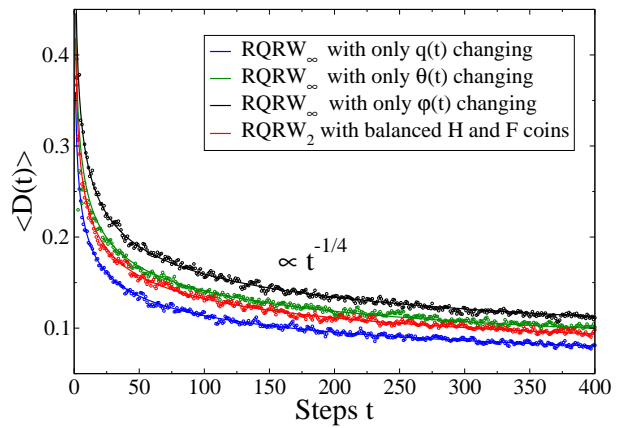}
\caption{\label{figdistancia2}(color online)
The curves were plotted averaging 
over 500 random initial conditions as given in Fig.~\ref{figdistancia}.
We worked with a $400$-step walk. The small circles are the simulated 
$\langle D(t)\rangle$ and the solid lines fitted curves to the data.
We assumed that $q(t)\in [0,1], \theta(t)\in [0,\pi]$, and $\varphi(t)\in [0,2\pi]$.}
\end{figure}

We have also studied the behavior of $\langle D(t)\rangle$ for the other 
$RQRW_{\infty}$ and for $RQRW_2$. As can be seen in Fig.~\ref{figdistancia2}, 
for all cases $D(t)\sim t^{-1/4}$.

Finally, we have extended the previous analysis to the matrix coefficients
of $\rho_C(t)$, i.e., we have computed $\langle |\alpha(t) - \alpha(t-1)|\rangle$, 
$\langle |\mathrm{Re}[\gamma(t)] - \mathrm{Re}[\gamma(t-1)]|\rangle$, and
$\langle |\mathrm{Im}[\gamma(t)] - \mathrm{Im}[\gamma(t-1)]|\rangle$
for several hundreds steps and initial conditions.
For the Hadamard QRW the dominant quantities go to zero as $t^{-1/2}$
and for all RQRW as $t^{-1/4}$. These and the previous results are clear
numerical indications that  $\rho_C(t)$ has an asymptotic limit whose origin 
can be traced to the dynamics of those quantum random walks and that 
this limit is approached differently whether
we have an ordered or disordered walk.

\subsection{The importance of being random in time}

One may ask whether the origin of the maximal entanglement generation capability of RQRW
is in the randomness of the coin or in the fact that it
is time dependent. As we show below, its origin can be traced back to both ingredients 
occurring at the same time.

First, one calculation in Ref. \cite{cha12} shows that static disorder may decrease the effectiveness of 
a quantum walk to generate entanglement, far below the entanglement generation 
capacity of the ordered case.
In other words, the presence of random coins at each position/site of the walk, 
and fixed throughout the whole evolution 
(static disorder), does not help in the generation of entanglement. However, care should be taken
in generalizing this interpretation since in \cite{cha12} only one initial condition for a particular
type of static disorder was numerically investigated, and for just a 100-step walk.  

Second, what about the time-dependence alone? That is, what will be the behavior of the 
entanglement generation capacity of QRW if non-random time dependent coins are
employed? In what follows we investigate this matter for periodic time dependent
coins.
\begin{figure}[!ht]
\includegraphics[width=8cm]{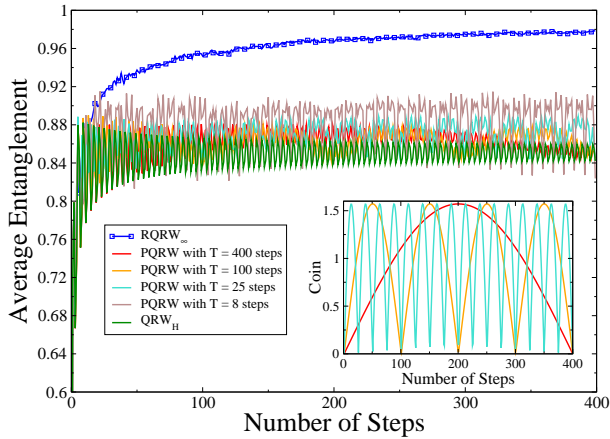}
\caption{\label{figper}(color online)
The curves were plotted averaging 
over 500 random initial conditions as given in Fig.~\ref{figdistancia}.
We worked with a $400$-step walk. The blue/square curve
represents $RQRW_\infty$ with $q(t)=1/2$ and $\theta(t)=\varphi(t)$
randomly chosen from a uniform distribution ranging from $0$ to $2\pi$.
The green curve represents the ordered QRW with the Hadamard coin.
The other curves are PQRW's evolving according to Eq.~(\ref{per1}). 
The inset shows how the coin ($\theta(t)$) changes during the whole evolution
for the PQRW's with $T=400, 100$, and $25$ steps.}
\end{figure}

In Fig. \ref{figper} we work with QRW's such that the coin oscillates 
smoothly (in a time-discretized sense) between the Hadamard and Fourier coins.
Therefore, for these coins $q(t)=1/2$ and $\varphi(t)=\theta(t)$, with
\begin{equation}
\theta(t) = \frac{\pi}{2}\left|\sin(\pi t/T)\right|.
\label{per1}
\end{equation}
Here $T$ is the period of oscillation and $t$ denotes the time-steps.
We call these walks periodic quantum random walks (PQRW).

\begin{figure}[!ht]
\includegraphics[width=8cm]{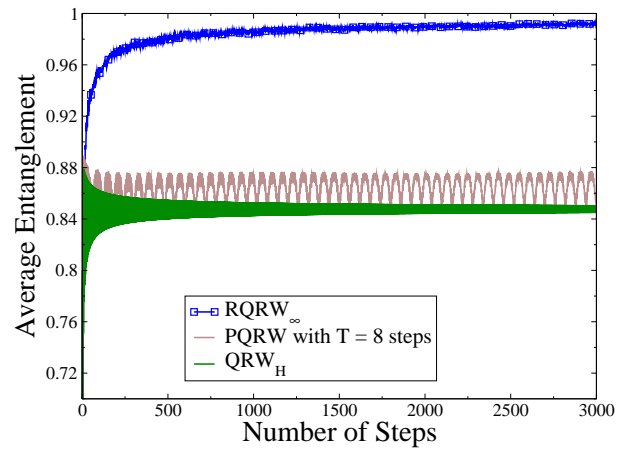}
\caption{\label{figlong}(color online)
The curves were plotted averaging 
over 200 random initial conditions as given in Fig.~\ref{figdistancia}.
We worked with a $3000$-step walk. The blue/square curve represents
the $RQRW_\infty$ as described in Fig.~\ref{figper}, 
the middle/brown curve $PQRW$ with $T=8$ steps and coin evolving 
according to Eq.~(\ref{per1}), and the lower/green curve represents 
the ordered QRW with the Hadamard coin.}
\end{figure}

As can be seen from Fig. \ref{figper}, only $RQRW_\infty$ approaches unity
entanglement asymptotically. Also, the average entanglement for all PQRW oscillates
with increasing amplitude as we decrease the period $T$.
We can also note that the oscillation of the average entanglement of 
the ordered case diminishes much faster than those for PQRW. Actually, 
for small $T$, the amplitude of oscillation for the average entanglement
decreases and then increases again with time for PQRW. 
This suggests that an asymptotic limit may not be 
achieved for such time-dependent coins. Or, if it occurs, it
will happen at very long times 
when compared to the ordered case
(see the brown/$T=8$ curve in Fig. \ref{figper}
and also Fig. \ref{figlong}).

%

\begin{figure}[!ht]
\includegraphics[width=8cm]{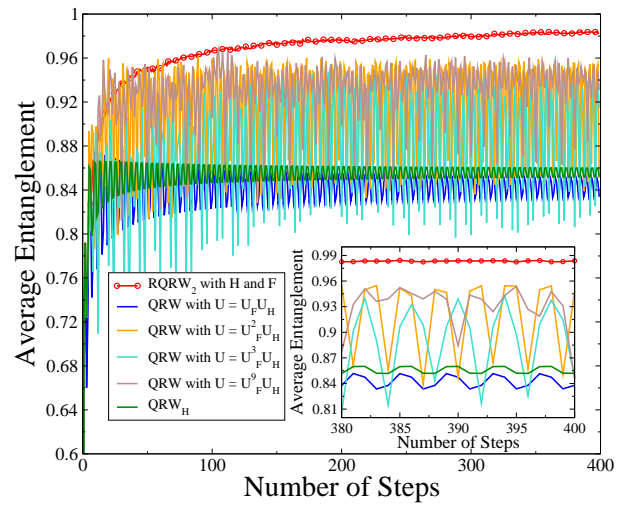}
\caption{\label{figper2}(color online)
The curves were plotted averaging 
over 500 random initial conditions as given in Fig.~\ref{figdistancia}.
We worked with a $400$-step walk.
The red/circle curve represents $RQRW_2$ with the Hadamard and Fourier coins.
The green curve represents the ordered QRW with the Hadamard coin.
The other curves are PQRW's evolving according to Eq.~(\ref{per2}). 
The inset shows the long time behavior of the curves given in the
main graph.}
\end{figure}

We have also investigated a less smooth time dependence, 
such that one oscillates directly between the Hadamard and the
Fourier coins. Noting that at each time-step the unitary evolution
acting on the walker is $U(t) = S(C(t)\otimes \mathbb{1}_P)$, 
we choose to work with the following time-dependent process
\begin{eqnarray}
|\Psi(t+T)\rangle &=& U_F(t+T)\cdots U_F(t+2)U_H(t+1)|\Psi(t)\rangle 
\nonumber \\ 
&=& U_F^{T-1}U_H|\Psi(t)\rangle,
\label{per2}
\end{eqnarray}
where $U_H(t)=U_H=S(H\otimes \mathbb{1}_P)$ and
$U_F(t)=U_F=S(F\otimes \mathbb{1}_P)$. Here $H$ and
$F$ are the matrices representing, respectively, the
Hadamard and Fourier coins and $T$ is the period, after
which the whole sequence of unitaries repeats 
itself. 

In Fig. \ref{figper2} we show the behavior of
the average entanglement for several periods $T$.
We note that the average entanglement oscillates
with greater amplitudes as compared to the smooth case. 
And as before, only the average entanglement for $RQRW_2$ clearly 
approaches the maximal value possible as we increase the number of steps. 

Although more systematic studies are needed along these lines, in particular
for static disorder, the results here presented suggest that neither
time independent (static) disorder nor non-random time dependent coins are sufficient to
generate maximally entangled states for any initial condition. 
Rather, random time dependent coins (dynamic 
disorder) appears to be essential to achieve such a feat.

\section{More numerical results}
\label{apB}

\subsection{Several random initial conditions}

Here we give more details about other RQRW's.
In addition to the average entanglement $\langle S_E \rangle$ we show
the rate at which highly entangled states are generated, 
the average probability distribution for finding
the qubit at a given position, and
the dispersion for these RQRW's.

\begin{figure}[!ht]
\includegraphics[width=8cm]{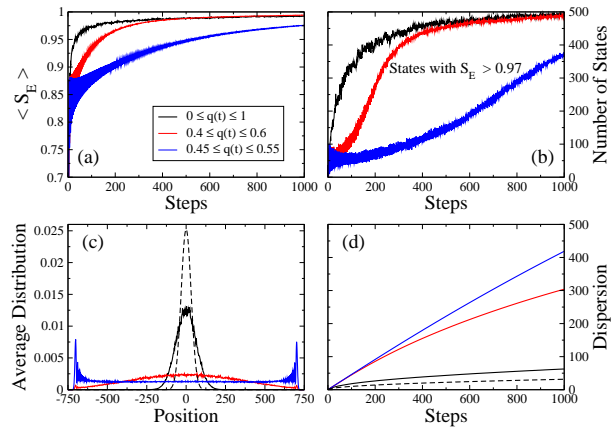}
\caption{\label{figAp1}(color online) The following graphs were plotted averaging 
over 500 random initial conditions of the following form,
$|\Psi(0)\rangle\hspace{-.1cm}=\hspace{-.1cm}(\cos\alpha_s|\hspace{-.1cm}\uparrow\rangle+e^{i\beta_s}
\hspace{-.1cm}\sin\alpha_s|\hspace{-.1cm}\downarrow\rangle) \otimes 
(\cos\alpha_p|-1\rangle+e^{i\beta_p}\hspace{-.1cm}\sin\alpha_p|1\rangle)$, 
where $\alpha_{s,p}\in [0,\pi]$ and $\beta_{s,p}\in [0,2\pi]$.  We worked with a $1000$-step walk
and with $RQRW_{\infty}$ where $\theta(t)=\varphi(t)=0$ and at each step  $q(t)$ was picked randomly from the three different 
uniform distributions given in (a), where the average entanglement $\langle S_E \rangle$ is shown for each
case. The other graphs are: (b) The number of states with $S_E > 0.97$ as a function of the
discrete time steps. (c) The average probability distribution $\langle P(j)\rangle$ after $1000$ steps, 
with the dashed line showing the classical case.
 (d) The average dispersion $\langle \sqrt{\sigma^2}\rangle$
at each step, with the dashed line showing the classical dispersion curve.
The classical case corresponds to a balanced CRW starting at the origin.}
\end{figure}

\begin{figure}[!ht]
\includegraphics[width=8cm]{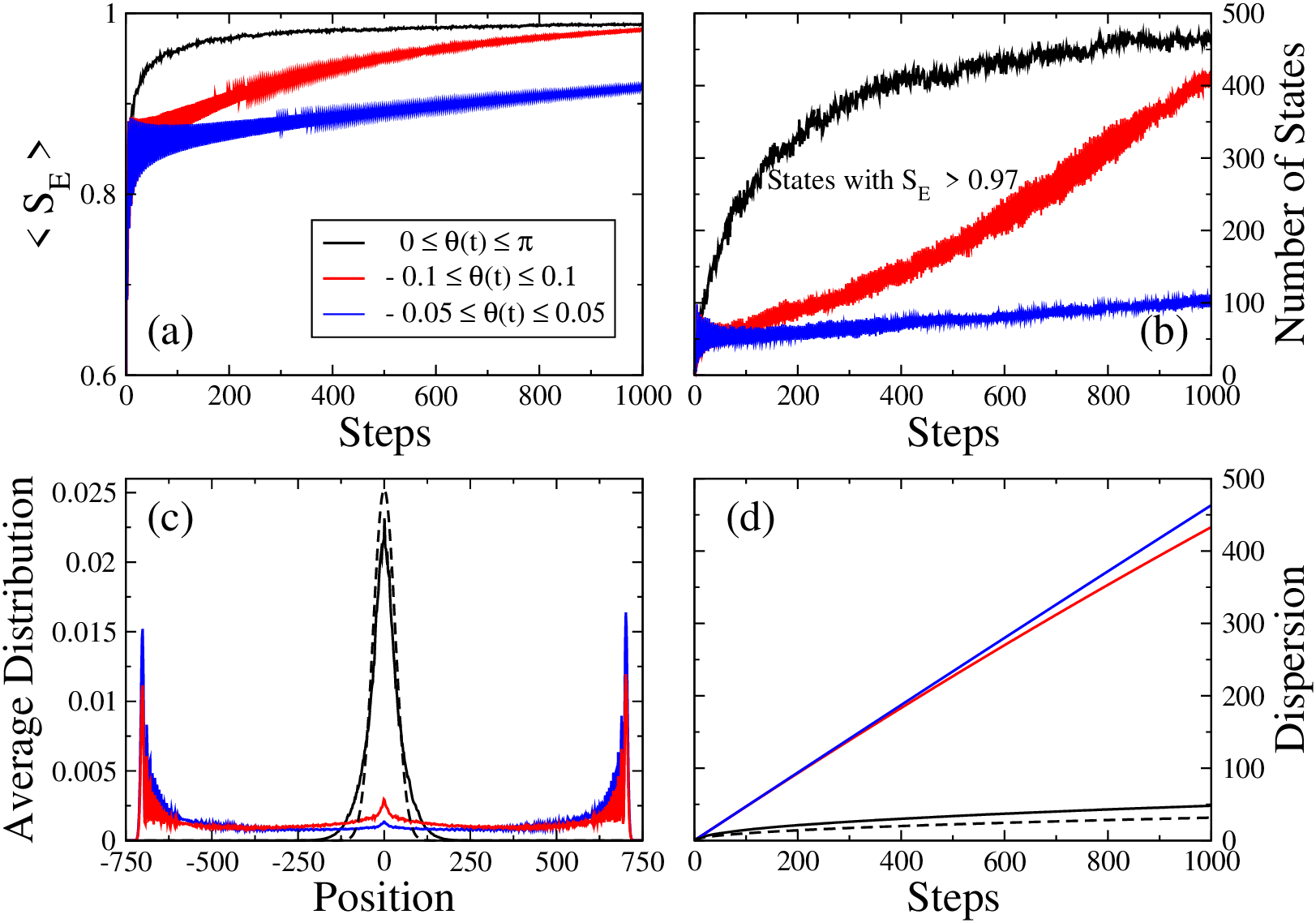}
\caption{\label{figAp2}(color online) The following graphs were plotted averaging 
over 500 random initial conditions as given in Fig.~\ref{figAp1} for a $1000$-step walk.
Here $RQRW_{\infty}$ was such that $q(t)=1/2$, $\varphi(t)=0$ and at each step $\theta(t)$ 
was randomly chosen from the three different 
uniform distributions given in panel (a), where the average entanglement 
$\langle S_E \rangle$ is shown for each
case. In panels (b), (c), and (d) we have the same as explained in Fig. \ref{figAp1}.
}
\end{figure}

\begin{figure}[!ht]
\includegraphics[width=8cm]{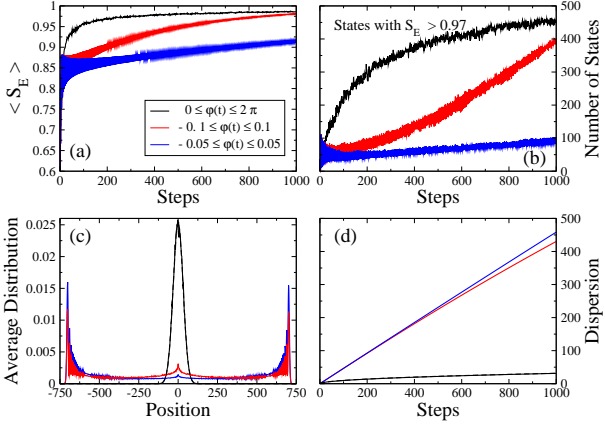}
\caption{\label{figAp3}(color online) The following graphs were plotted averaging 
over 500 random initial conditions as given in Fig.~\ref{figAp1} for a $1000$-step walk.
Here $RQRW_{\infty}$ was such that $q(t)=1/2$, $\theta(t)=0$ and at each step $\varphi(t)$ 
was randomly chosen from the three different 
uniform distributions given in panel (a), where the average entanglement 
$\langle S_E \rangle$ is shown for each
case. In panels (b), (c), and (d) we have the same quantities as given in Fig. \ref{figAp1}.
Note that here the classical case (dashed curves) overlap the solid/black ones.
}
\end{figure}

\begin{figure}[!ht]
\includegraphics[width=8cm]{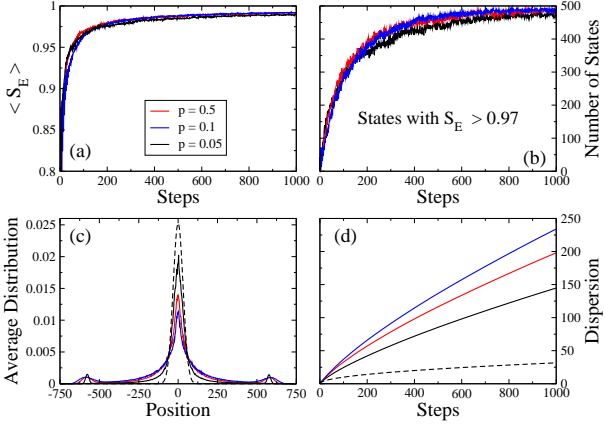}
\caption{\label{figAp4}(color online) The following graphs were plotted averaging 
over 500 random initial conditions as given in Fig.~\ref{figAp1} for a $1000$-step walk.
Here we implement $RQRW_2$ with $p$ being the probability of using at step $t$
the Hadamard coin and $(1-p)$ the Fourier coin. In panel (a) we see the average entanglement 
$\langle S_E \rangle$ for each value of $p$. 
The other panels show the same quantities as explained in Fig. \ref{figAp1}.
Note that weak disorder is enough
to guarantee $S_E\rightarrow 1$ asymptotically and it is remarkable
that a mild deviation ($p=0.05$) from the Fourier/Kempe $QRW$ leads
all initial conditions to almost perfectly entangled states after $1000$ steps.
This illustrate the efficiency of this process
to generate highly entangled states, even when weak disorder is present ($p=0.05$).}
\end{figure}

The first thing worth mentioning is the robustness of these RQRW's to generate highly entangled states, even
if we have random coins very close to the Hadamard coin, i.e, even for not too big deviations from the values of 
$q(t)=1/2$ and $\theta(t)=\varphi(t)=0$ that characterize the H coin.  Also, $RQRW_{\infty}$ with
only $q(t)$ random (panel (b) of
Fig. \ref{figAp1}) has higher generation rates of highly entangled states than the other two $RQRW_{\infty}$,
with either only $\theta(t)$ or $\varphi(t)$ changing (see panels (b) of Figs. \ref{figAp2} and \ref{figAp3}). 
And if we look at Fig. \ref{figAp4}.b, we see
that $RQRW_2$ with H and F coins is the most efficient entanglement generator. 
It can produce highly entangled states for almost any initial condition within just a few hundreds
steps, even for a very tiny deviation from the Fourier QRW ($p=0.05$).

We also note that the average probability distribution $\langle P(j)\rangle$ and the dispersion possess
a wide range of behaviors. However, there is a common trend for all RQRW's: 
the more we deviate from a fixed coin, the more it approaches
the classical case. Also, among all $RQRW_{\infty}$ cases, 
the one in Fig.~\ref{figAp1} gives the smallest peaks far from the origin (Fig.~\ref{figAp1}.c)
while the other two cases exhibit the greatest symmetric peaks away from it. Furthermore, the
weaker the disorder the greater the peaks. This is expected since for weaker disorder we
are approaching the fixed coin case, where these peaks are a common trend.

\subsection{Delocalized Gaussian initial conditions}

We now study the ability of RQRW's against the standard QRW
to entangle the external and internal 
degrees of freedoms of the walker for highly non-local/delocalized
initial conditions in position. We will work with $RQRW_{\infty}$, with $q(t)$, $\theta(t)$, and $\varphi(t)$ completely
random, and the Hadamard QRW.

\begin{figure}[!ht]
\includegraphics[width=8cm]{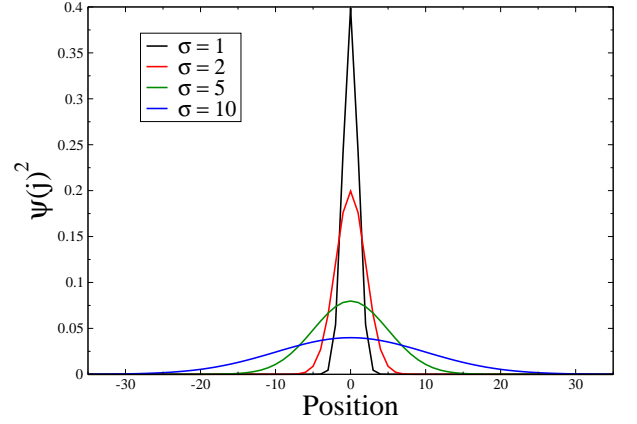}
\caption{\label{gauss}(color online) The wider the Gaussian curve the greater 
$\sigma$.}
\end{figure}

The several initial positions of the qubit are given by
Gaussian distributions centered about the origin, 
$$
\psi^2(x)=\frac{1}{\sqrt{2\pi\sigma^2}}e^\frac{-x^2}{2\sigma^2},
$$
where $\sigma^2$ is the variance. The global initial state is therefore,
$$
|\Psi(0)\rangle=|\xi\rangle\otimes \sum_{j=-\infty}^{\infty}\psi(j)|j\rangle,
$$
where $\psi(j)\geq 0$ and we work with the discretized and normalized version of the 
Gaussian distribution.
In what follows we will be working with two types of initial spin states,
$|\xi_1\rangle=(|\uparrow\rangle$ $+i|\downarrow\rangle)/\sqrt{2}$ and 
$|\xi_2\rangle=(|\uparrow\rangle+|\downarrow\rangle)/\sqrt{2}$, and several Gaussians
with dispersions ($\sigma$) as given in Fig.~\ref{gauss}.

\begin{figure}[!ht]
\includegraphics[width=8cm]{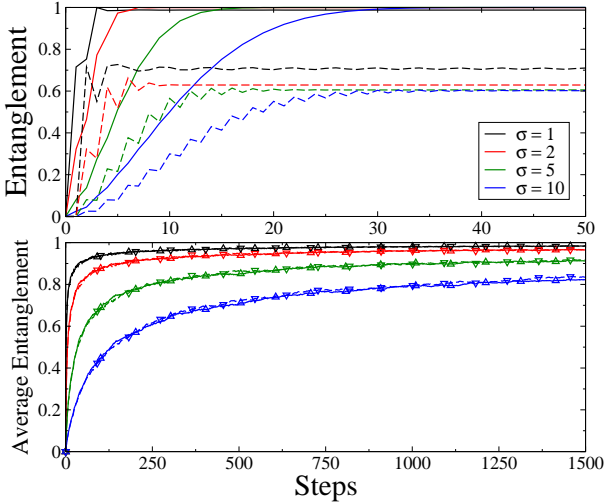}
\caption{\label{gaussSE}(color online) Upper panel: The solid curves represent the entanglement for 
Gaussian initial conditions with spin state $|\xi_1\rangle$ and the dashed ones with
$|\xi_2\rangle$. We are working with the Hadamard QRW. Note that the asymptotic entanglement
is extremely sensitive to initial conditions, where we can see it approaching a wide range of 
possible values. Lower panel: The same initial conditions as above but now we work with
$RQRW_{\infty}$ where $q(t)\in [0,1], \theta(t)\in [0,\pi]$, and $\varphi(t)\in [0,2\pi]$. 
The solid and dashed curves are almost the same, so we 
employ up triangles to the cases starting with  $|\xi_1\rangle$ and down triangles to 
those starting with  $|\xi_2\rangle$. We now see that all cases approach
the maximal entanglement value ($S_E \rightarrow 1$), although the greater the initial dispersion $\sigma$ 
the slower the rate at each the maximal value is approached. 
For $RQRW_{\infty}$ we implemented
$500$ different realizations and plotted the average entanglement.}
\end{figure}

\begin{figure}[!ht]
\includegraphics[width=8cm]{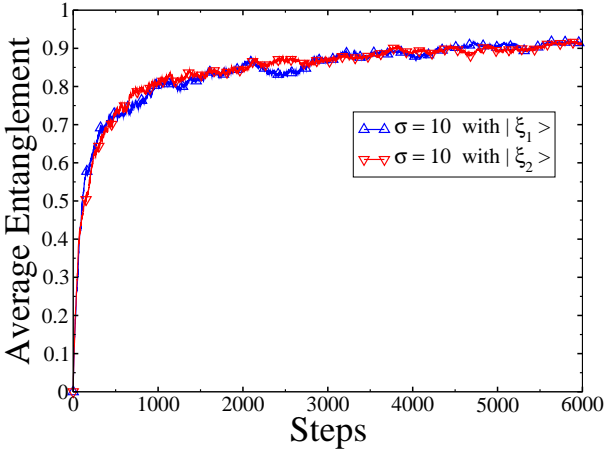}
\caption{\label{6000}(color online) We implement $100$ realizations of 
$RQRW_{\infty}$ with $q(t)\in [0,1], \theta(t)\in [0,\pi]$, and $\varphi(t)\in [0,2\pi]$. 
We go up to $6000$ steps. Up triangles refer to $\langle S_E \rangle$ for the
cases starting with  
$|\xi_1\rangle$ and down triangles to $|\xi_2\rangle$. Now we obtain
$\langle S_E\rangle > 0.9$ for a Gaussian with $\sigma=10$.}
\end{figure}

In Fig. \ref{gaussSE} we plot the entanglement for both walks starting with the previous
initial conditions. It is clear that the asymptotic entanglement of the Hadamard QRW is highly sensitive to
initial conditions and many do not lead to maximal entanglement. For $RQRW_{\infty}$, we see the independence on
the initial state for the asymptotic value of entanglement, although the rate at which it 
is approached depends on the broadness of the initial probability distribution for the position
of the particle. Indeed, as we show in Fig.~\ref{6000}, for $\sigma = 10$ the walker needs to 
travel about $6000$ steps to achieve $\langle S_E \rangle > 0.9$. For just $1500$ steps,
we have $\langle S_E \rangle \approx 0.8$. 

\begin{figure}[!ht]
\includegraphics[width=8cm]{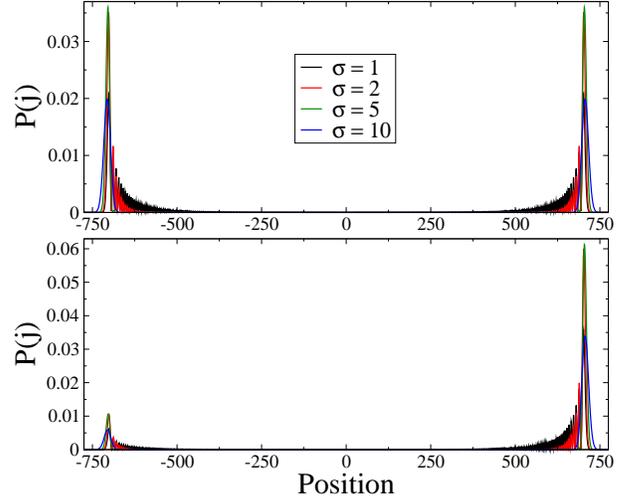}
\caption{\label{gaussQRWH}(color online) The upper panel shows the probability distributions 
$P(j)$ after $1500$ steps for initial conditions given by Gaussians with dispersion
$\sigma$ and initial spin state $|\xi_1\rangle$ while the lower panel shows the cases with
spin state $|\xi_2\rangle$.
In both cases we have the Hadamard QRW. Note that $|\xi_1\rangle$ produces symmetrical peaks while
$|\xi_2\rangle$ asymmetrical ones. The wider the initial conditions the wider the peaks.}
\end{figure}

\begin{figure}[!ht]
\includegraphics[width=8cm]{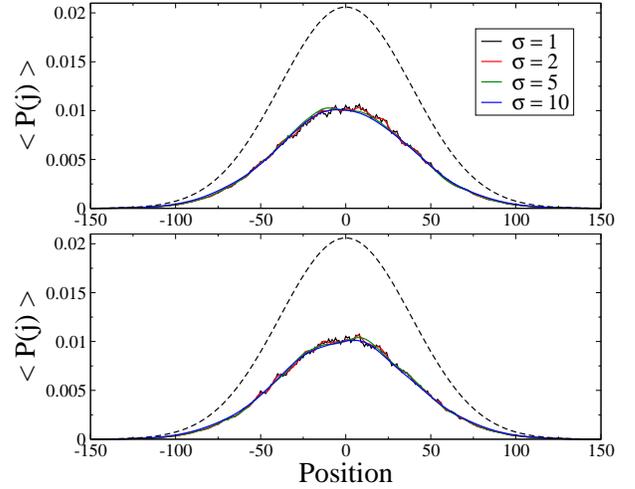}
\caption{\label{gaussRQRWInf}(color online) The upper panel shows the average probability distributions 
$\langle P(j)\rangle$ over $500$ realizations after $1500$ steps for initial conditions given by Gaussians 
with dispersion $\sigma$ and initial spin state $|\xi_1\rangle$. The lower panel shows the cases with
spin state $|\xi_2\rangle$.
In both cases we have $RQRW_{\infty}$ where $q(t)\in [0,1], \theta(t)\in [0,\pi]$, and $\varphi(t)\in [0,2\pi]$. 
Now all spin initial conditions 
give similar centralized peaks. The dashed line shows the expected classical $P(j)$
for a balanced CRW starting at the origin.}
\end{figure}

Finally, in Figs. \ref{gaussQRWH} and \ref{gaussRQRWInf} we show the probability distribution after 
$1500$ steps for walkers starting with the initial conditions given in 
Fig.~\ref{gauss} for the Hadamard
QRW and  $RQRW_{\infty}$, respectively.

\section{Experimental predictions}
\label{apC}

The two experiments \cite{sch10,cre13} on which our experimental proposal is built achieve 
so far a few tens of steps. In particular, in \cite{sch10} RQRW's with 28 steps were
implemented. Our goal here is, therefore, to show that with just tens of steps we already have
different predictions for the entanglement $S_E$ whether we implement the standard QRW or RQRW. 

To build our experimental proposal, we started with the localized initial condition 
$|\Psi(0)\rangle=(\cos\alpha_s|\hspace{-.1cm}\uparrow\rangle+e^{i\beta_s}
\hspace{-.1cm}\sin\alpha_s|\hspace{-.1cm}\downarrow\rangle)$ $\otimes 
|0\rangle$ and searched in increments of $0.1$ 
for the pair of points $(\alpha_s,\beta_s)$ giving the lowest $S_E$
for the Hadamard QRW at step $28$. We found $(\alpha_s,\beta_s)=(2.7,\pi)$ with
$S_E = 0.645$. By keeping $\alpha_s=2.7$ and changing $\beta_s$ we get several
different values for $S_E$ at step $28$ for the Hadamard QRW: 
$(\alpha_s,\beta_s)=(2.7,\pm \pi)\rightarrow S_E=0.645$, 
$(\alpha_s,\beta_s)=(2.7,0)\rightarrow S_E=0.983$, and
$(\alpha_s,\beta_s)=(2.7,\pm \pi/2)\rightarrow S_E=0.869$. All these initial
conditions can be easily prepared with half (HWP) and quarter wave plates (QWP).

Then, with the initial condition giving the lowest $S_E$ for the Hadamard QRW, we implemented
$10,000$ numerical experiments using the balanced $RQRW_2$ with the
Hadamard (H) and Fourier/Kempe (F) coins,  searching for the sequence of random 
H's and F's giving the greatest entanglement. Using this sequence, we computed the 
entanglement for all initial conditions above:
 $(\alpha_s,\beta_s)=(2.7,\pm \pi)\rightarrow S_E=0.999$, 
$(\alpha_s,\beta_s)=(2.7,0)\rightarrow S_E=0.979$,
$(\alpha_s,\beta_s)=(2.7, \pi/2)\rightarrow S_E=0.938$, and
$(\alpha_s,\beta_s)=(2.7, -\pi/2)\rightarrow S_E=0.982$. As can be seen, with
the $RQRW_2$ we already have after $28$ steps all these five initial conditions
giving $S_E > 0.93$, with four of them giving $S_E > 0.97$. In contrast, the 
Hadamard QRW has only one initial condition giving $S_E>0.93$ and two of them
giving $S_E < 0.65$. This huge contrast can be experimentally detected with current
day technology.

The sequence of H's and F's leading to such predictions is (with time flowing
from left to right),
$$
HHFHFFHFHFFHFHFHFFHFFHHFHFHH,
$$
and it can be implemented in \cite{sch10} by adjusting the phase-shifter before
the passage of the photon to the HWP that generates the standard coin and
in \cite{cre13} by writing the integrated waveguide circuit with the correct
optical path differences between successive
directional couplers (the equivalent of polarizing beam splitters).

\begin{figure}[!ht]
\includegraphics[width=8cm]{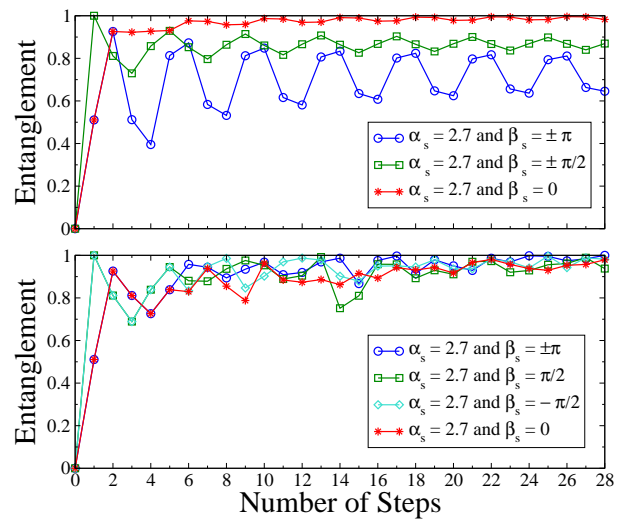}
\caption{\label{figPrediction}(color online) In the top panel we have the entanglement evolution
for the Hadamard QRW and in the bottom panel its evolution for the $RQRW_2$ with a sequence of
H and F coins as given in the text. 
Note that the entanglement at step $28$ for all initial conditions clusters around $0.9$ for 
$RQRW_2$ while it appreciably differs for different initial conditions for the Hadamard QRW.
Also, note that for QRW the entanglement evolves oscillating periodically about its
asymptotic value. For $RQRW_2$ this periodicity is lost.}
\end{figure}

In Fig.~\ref{figPrediction} we show
the entanglement time evolution for both
the Hadamard $QRW$ and $RQRW_2$ for all the previous five initial conditions.

The same analysis can be carried out to other RQRW's in order to find a sequence of random
coins that clearly gives different predictions for RQRW and QRW within just a few steps. In the
asymptotic limit, of course, any random sequence will do.


\begin{thebibliography}{200}
\bibitem{mudanca1} This internal degree of freedom can be the spin of an electron,
the polarization of photons or the ground and excited states of an atom.
\bibitem{application} CRW is employed from the
modeling of biological processes \cite{ber93} to polymer physics
\cite{gen79}. QRW has proved to give
important insights into the implementation of quantum search
algorithms \cite{she03}, the understanding of the physics of
photosynthesis \cite{eng07}, the construction of a universal quantum
computer \cite{chi09,lov10}, and the generation of entangled states for
systems with one \cite{car05} and more walkers
\cite{bos09}.
\bibitem{experiment1} The random coin operator $C$ is implemented using
HWP and EOM while the conditional displacement operator $S$
two PBS's and a fiber delay line where one polarization
follows a longer optical path. Also, arbitrary initial conditions are simply generated by
QWP and HWP. An important feature of the scheme in
\cite{sch10} is its scalability with the number of steps. Indeed,
the techniques of optical feedback loop allow the implementation of a
many-step walk using few optical elements. In
\cite{sch10} the authors
have already implemented $28$-step walks with static and dynamical disorder.
\bibitem{experiment2} In integrated
waveguide circuits PBS is a directional coupler and
phase shifts are implemented writing circuits with different
length/deformation.
\bibitem{footnote3} The numerical results for the long time behavior in what follows are unchanged
if we assume for example $0\leq \theta(t)\leq\pi$ and $0\leq\varphi(t)\leq 2\pi$. 
Also, the analytical proof of the theorem does not rely on specific choices for the ranges of  
$\theta(t)$ and $\varphi(t)$.
\end{thebibliography}
\end{document}